\documentstyle[prb,multicol,aps,graphicx]{revtex}
\begin{document}
\draft 
\title{Characterization of transport and magnetic properties 
in thin film 
La$_{0.67}$(Ca$_{x}$Sr$_{1-x}$)$_{0.33}$MnO$_{3}$ mixtures} 
\author{P. R. Broussard, S.B. Qadri, V.M. Browning, and V. C. Cestone} 
\address{Naval Research Lab, 
Washington, DC 20375} 
\bigskip 
\maketitle
\begin{abstract}
We have grown thin films of (100) oriented 
La$_{0.67}$(Ca$_{x}$Sr$_{1-x}$)$_{0.33}$MnO$_{3}$ on (100) NdGaO$_{3}$ 
substrates by off-axis sputtering.  We have looked at the changes in 
the resistivity and magnetoresistance of the samples as the Ca/Sr 
ratio was varied. We find that as the calcium fraction is decreased, 
the lattice match to the substrate decreases, and the films become 
more disordered, as observed in transport measurements and the 
variation in Curie and peak resistance temperatures.  We find a 
correlation between the temperature independent and T$^{2}$ terms to 
the low temperature resistivity.   The room 
temperature magnetoresistance exhibits a maximum as the peak temperature is increased by 
the substitution of Sr for Ca, and a change in the field 
dependence to the resistivity at room temperature is observed.
\end{abstract}
\bigskip


\begin{multicols}{2}
	
\narrowtext

\section{Introduction}
The effect of dopants for the ABO$_{3}$-type manganese oxides has been 
an area of intense research activity, primarily for attempting to 
understand the physics behind the colossal magnetoresistance (CMR) 
behavior seen in these materials.  Typically, studies have been 
carried out either by replacing the trivalent 
ion\cite{Sun,Fontcuberta,Hwang} or the Mn 
ion.\cite{Blasco,Gayathri,Anane} Recently\cite{Guo} a study of 
polycrystalline La$_{0.75}$Ca$_{0.25-x}$Sr$_{x}$MnO$_{3}$ was carried 
out in order to look at changes in magnetic entropy, where the doping 
variation is on the divalent site.  In the work by Hwang et 
al.\cite{Hwang} the system 
La$_{0.7}$(Ca$_{x}$Sr$_{1-x}$)$_{0.3}$MnO$_{3}$ was looked at, and 
exhibited a change in the tolerance factor t, which is defined as 
$t=(d_{A-O})/\sqrt{2}(d_{Mn-O})$, from $\approx$ 1.2 to 1.24.  
We have undertaken a study of the 
changes in electronic, structural, and magnetic properties of thin 
films of La$_{0.67}$(Ca$_{x}$Sr$_{1-x}$)$_{0.33}$MnO$_{3}$ (LCSMO) as the 
Ca/Sr ratio is varied.  

\section{Sample preparation and characterization}
Our samples were grown by off-axis sputtering using composite targets 
of La$_{0.67}$Ca$_{0.33}$MnO$_{3}$ (LCMO) and 
La$_{0.67}$Sr$_{0.33}$MnO$_{3}$ (LSMO) material mounted in copper 
cups.  The substrates were (100) oriented neodymium gallate 
(NdGaO$_{3}$), silver-pasted onto a stainless steel substrate holder 
that was radiatively heated from behind by quartz lamps.  Although 
there was no direct measurement of the holder temperature for the runs 
used in this study, previous runs (under nominally the same 
conditions) using a thermocouple clamped onto the front surface of the 
holder indicated a temperature of 670 C. The LCMO target was 
radio frequency (rf) sputtered and the LSMO target was direct current 
(dc) sputtered in a sputter gas 
composed of 80\% Ar and 20\% O$_{2}$ (as measured by flow meters) and 
at a total pressure of 13.3 Pa.  These conditions gave deposition 
rates of $\approx$ 17-50 nm/hr, with film thicknesses being typically 
100 nm.  After deposition, the samples were cooled in 13.3 kPa of 
oxygen.  We find that our system can produce films of LCMO and LSMO 
that have low resistivities and high peak temperatures without the use 
of an {\it ex-situ} anneal in oxygen.  

The samples were characterized by standard and high resolution $\theta-2\theta$ x-ray 
diffraction scans, atomic force microscopy, electrical resistivity 
measurements (using the van der Pauw method\cite{VdP}) in an applied 
field perpendicular to the film plane, and magnetization measurements 
at low fields parallel to the film plane using a Quantum Design SQUID 
Magnetometer.  All magnetization data had the large paramagnetism of 
the NdGaO$_{3}$ substrates subtracted out.

\section{Structure, Transport and Magnetic Properties}

\begin{figure}
\begin{center}
\includegraphics*[width=0.45\textwidth]{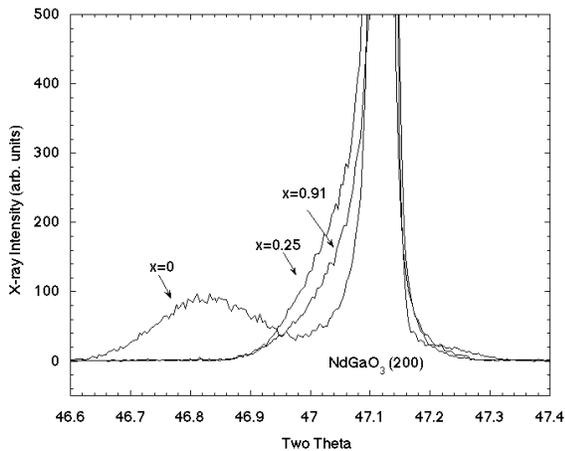}
\end{center}
\caption{High resolution X-ray diffraction scans along the film normal 
for a pure LSMO film (x=0) and two LCSMO mixtures (calcium fraction 
x=0.25 and x=0.91)
grown on NdGaO$_{3}$.}
\label{XRD}
\end{figure}

On (100) NdGaO$_{3}$ we find 
surface roughness values of $\approx$ 1.5 nm for pure films of LSMO 
and LCMO, while for the mixtures the surface roughness increases to 
$\approx$ 2.8 nm, as measured by atomic force 
microscopy.  The grain sizes for the pure LSMO and LCMO films is 
typically 100 nm, while for the mixtures it is reduced to $\approx$ 50 
nm.  High resolution X-ray diffraction along the growth direction 
shows only the presence of peaks from NdGaO$_{3}$ for the LCMO samples.  This would be 
expected, since the lattice match of pseudo-cubic (100) LCMO 
($a_{o}$ $\approx$ 0.387 nm) to pseudo-cubic (100) NdGaO$_{3}$ ($a_{o}$ $\approx$ 
0.385 nm) should be excellent.  From this we take the orientation 
of the LCSMO films to be (100).  Films of LSMO on NdGaO$_{3}$ however 
as shown in Fig. \ref{XRD} do exhibit a peak corresponding to a 
pseudo-cubic length of 0.388 nm.  The rocking curve width for this 
line is 337 arc-seconds, with an instrumental width of 12 
arc-seconds, and phi scans show excellent in-plane registry of the film with 
the NdGaO$_{3}$ substrate.  From our work on LSMO and LCMO grown on both (100) and 
(110) MgO, we know that LSMO films typically grow with a slightly 
larger value of the pseudo-cubic cell length, $a_{o}$, compared to LCMO.  On 
(100) MgO, for example, we find that $a_{o}$ is 0.387 and 0.388 nm for LCMO and 
LSMO, respectively, while for (110) MgO we find 0.388 and 0.390 nm 
for the two materials.  Obviously the lattice match is not as good for 
the case of LSMO, and this will introduce strain into the LSMO film.  
We also see from Fig. \ref{XRD} that as the calcium fraction is 
increased, the well defined peak seen for the LSMO film moves to 
smaller d spacings, consistent with the trend towards LCMO, appearing 
as a shoulder on the low angle side of the NdGaO$_{3}$ (200) peak.  As 
the calcium fraction increases further, the shoulder diminishes, and 
for pure LCMO (not shown), it is indistinguishable from the substrate 
peak.  From this we surmise that the films will be strained, with the 
strain decreasing as the calcium fraction increases.

\begin{figure}
\begin{center}
\includegraphics*[width=0.45\textwidth]{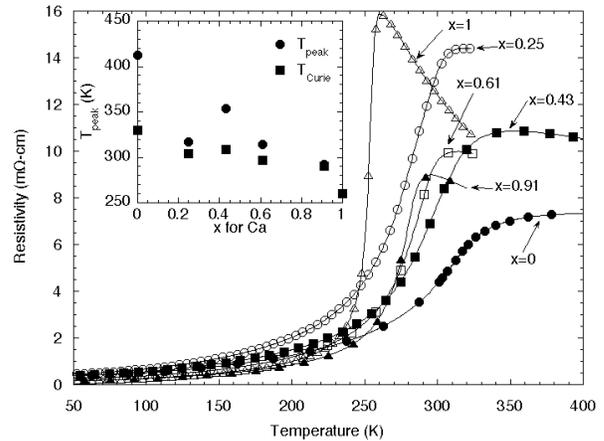}
\end{center}
\caption{Electrical resistivity vs. temperature for LCSMO films 
grown on (100) NdGaO$_{3}$ for different values of x, the calcium 
fraction.  The insert shows the variation in the peak temperature and 
Curie temperature as a function of the calcium concentration.}
\label{rho1}
\end{figure}

In Fig. \ref{rho1} we present the resistivity data in zero applied 
magnetic field for the LCSMO films for the various Ca/Sr ratios, along 
with a plot of the peak temperatures (T$_{p}$) and Curie 
temperatures (T$_{C}$) determined from magnetization data for the 
samples (taken at 400 Oe).  For the case of pure LCMO, we see the 
usual $\rho$(T) behavior, with a thermally activated resistivity 
(activation energy of $\approx$ 52 meV) and a peak temperature of 260 
K, which is the same as the Curie temperature.  For pure LSMO (x=0), 
we find that the resistivity has a peak temperature (410 K) much 
higher than the Curie temperature (330 K) This discrepancy between the 
peak and Curie temperatures is often seen for 
LSMO. The Curie temperature we see for our LSMO is lower than that 
seen in bulk LSMO with 1/3 doping,\cite{Urashibara} which we feel is 
due to disorder in the sample.  The difference in disorder or strain between 
the LSMO and LCMO samples was also seen in the measurements of the 
coercive field for the two samples.  At 10 K, the coercive field for 
the LCMO film was 20 Oe, which is quite low.  However for the LSMO 
film, the coercive field was 170 Oe.  Now as the concentration of Ca is 
varied from either extreme, we see sudden changes in the sample 
properties.  For the x=0.91 sample, we see a large decrease in the 
sample resistivity, with a concurrent rise in the T$_{p}$ and T$_{C}$.  
The increase in the Curie temperature is likely due to the change in 
the tolerance factor as the Ca atoms are replaced with Sr atoms.  On 
the other end, at x=0.25, the increase in Ca fraction causes a large 
increase in the resistivity, along with a large drop in T$_{p}$ and 
T$_{C}$.  The large increase in resistivity is likely the reason for 
the low value of T$_{C}$ observed.  The values for T$_{p}$ and T$_{C}$ 
are similar for the x=0.91 sample, but diverge for the samples with 
lower x.  We see that as the Ca fraction decreases, T$_{p}$ and 
T$_{C}$ tends to increase as one would expect for the change in 
tolerance factor being introduced.  However, the variation we see in 
T$_{C}$ is not as gradual as was seen in either the previous bulk 
studies on divalent doping.\cite{Hwang,Guo}  In the work by Hwang et 
al. an apparent smooth variation in T$_{C}$ from 250 to 365 K is seen as the Ca/Sr 
ratio is varied.  In the work by Guo et al. there was a jump in 
T$_{C}$ at a Ca fraction, x, of $\approx$ 0.45 when the system went 
from orthorhombic to rhombohedral.  No such jump is seen in our data, 
but with the strong lattice match to the NdGaO$_{3}$ substrate, we 
would not expect a structural change.  We see instead that the 
value of T$_{C}$ changes very slowly as the value of x is changed, 
but with rather sudden changes near x=0 and x=1.  We feel that part of the 
explanation for the non-monotonic behavior of T$_{C}$, as well as T$_{p}$, is due to 
disorder in the samples, as we discuss below.

\newpage

\begin{figure}
\begin{center}
\includegraphics*[width=0.45\textwidth]{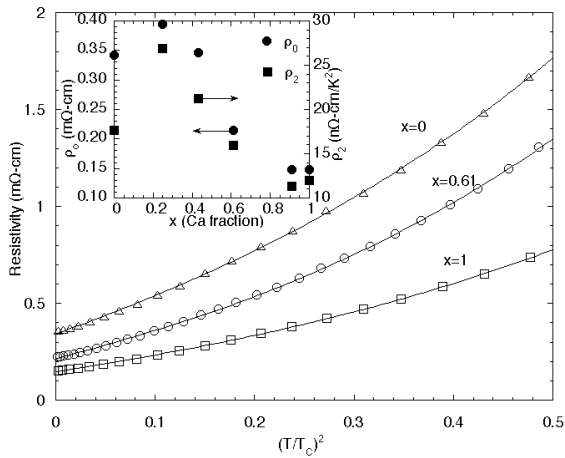}
\end{center}
\caption{Low temperature resistivity vs. (T/T$_{C}$)$^{2}$ for a LCMO, LSMO and x=0.61 
LCSMO film.  The insert shows the derived values of $\rho_{0}$ and 
$\rho_{2}$ vs. calcium fraction.}
\label{lowrho}
\end{figure}

In previous studies of the low temperature (T $<$ 200 K) resistivity 
in the manganites, several different equations have been used to 
characterize the behavior.  Schiffer et al.\cite{Schiffer} used the 
equation $\rho (T) = \rho_{0} + \rho_{1}T^{2.5}$ for LCMO 
polycrystalline material and found good fits to the data.  Similar 
results for LCMO films have also been seen.\cite{Vlakhov} Urashibara 
et al.\cite{Urashibara} found for LSMO material a T$^{2}$ dependence, 
which was interpreted as being due to electron-electron scattering.  
We found that we could also get reasonable fits using the approach in 
Schiffer et al.  if we limited the data selection to T $\rm <$ 150 K. 
However, if we look at the data for T $\rm <$ 200 K, we find that we 
get better agreement if we use
\begin{equation}
\rho (T) = \rho_{0} + \rho_{2}T^{2} +\rho_{5}T^{5}
\label{rhoT}
\end{equation}
as seen in Figure \ref{lowrho}.  A similar result was seen in well 
annealed LSMO and LCMO films by Snyder et al.,\cite{Snyder} but here 
they used a $T^{4.5}$ term instead of $T^{5}$, in light of the 
prediction of spin-wave scattering by Kubo and Ohata.\cite{Kubo} 
However, from the work on Pb-doped LCMO single crystals,\cite{Jaime} 
the contribution from the $T^{9/2}$ term is expected to be much 
smaller ( $\approx$ 0.5 $\mu \Omega$-cm at 100 K) than that seen in our results, 
which is $\approx$ 10 $\mu \Omega$-cm.  
We also observe the reduction in the contribution of the $T^{2}$ term 
at low temperatures, which is 
interpreted by Jaime et al. as an indication that the $T^{2}$ term 
arises from single-magnon scattering, and not electron-electron 
scattering.

\begin{figure}
\begin{center}
\includegraphics*[width=0.45\textwidth]{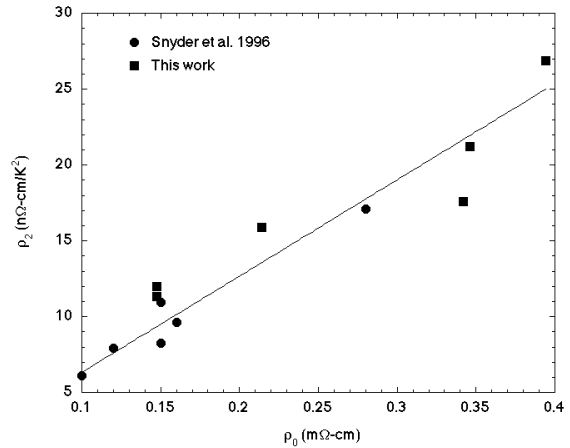}
\end{center}
\caption{Values of $\rho_{2}$ vs $\rho_{0}$ for this work and the thin 
films in Reference \cite{Snyder}.  The line is a guide to the eye.}
\label{rhocomp}
\end{figure}

Our derived values for $\rho_{0}$ and $\rho_{2}$ determined from 
fitting Eq.\ref{rhoT} are shown in the inset to Fig. \ref{lowrho}.  
The values of $\rho_{5}$ are typically 1 f$\Omega$-cm $K^{-5}$.  We see 
that the temperature independent term, $\rho_{0}$ is lowest for the 
pure LCMO films, with values similar to that seen in Snyder et 
al.\cite{Snyder} As the Ca fraction decreases, we see an increase in 
$\rho_{0}$, which indicates an increase in the disorder in the 
films.  This increase in disorder might be initially thought to be due 
to random location of Sr on Ca sites, but since the Ca sites are 
already located at random in LCMO, it is difficult to see how 
replacing Ca with Sr has increased the randomness in the system.  The trend continues until pure 
LSMO is reached, when we see a drop in the static term.  We notice 
however that the low temperature resistivity is higher for our pure 
LSMO films than for pure LCMO, which reflects the increased disorder 
for the LSMO film as seen in the coercive field measurements. A 
similar result was also seen in 
Ref.\cite{Snyder}.  For the temperature dependent term, we see a 
similar non-monotonic trend, with a peak in the value of $\rho_{2}$ as the Ca 
fraction decreases, and a large drop when pure LSMO is reached.  The 
values of $\rho_{2}$ that we observe for pure LSMO and LCMO are 
larger than that seen in Snyder et al.\cite{Snyder}, however both 
our values and those for Snyder show a similar correlation, as seen 
in Fig. \ref{rhocomp}.  Clearly there appears to be a connection between the 
values of $\rho_{2}$ and $\rho_{0}$, with the ratio of the two being 
approximately 60-70 x 10$^{-6}$ K$^{-2}$.  If the $\rho_{2}$ term is due to 
electron-electron scattering, it is very hard to see what correlation 
would exist between the static disorder in the sample and the terms in 
e-e scattering, such as the Fermi energy.  The model of Jaime et 
al.\cite{Jaime} also would give no correlation between the two 
terms.  If there was a coincidental correlation between the 
two terms, due say to changes in E$_{F}$ (which affects 
$\rho_{2}$) and changes in strain (which affects $\rho_{0}$) with x, 
we would not expect to see the same correlation for the films in the 
work by Snyder et al.\cite{Snyder} since the points with the lowest 
and highest values of $\rho_{0}$ are for pure LCMO films. 

\begin{figure}
\begin{center}
\includegraphics*[width=0.45\textwidth]{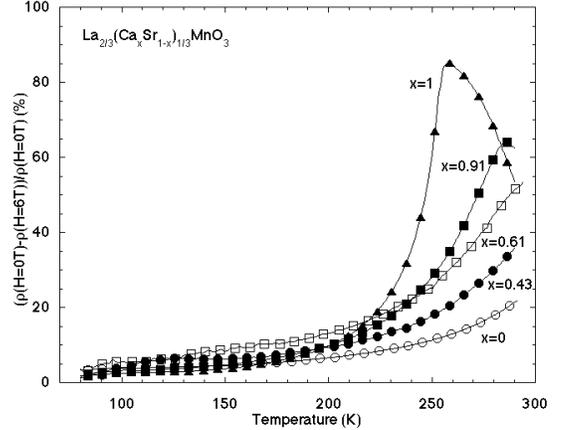}
\end{center}
\caption{Magnetoresistance vs. temperature at 6 Tesla applied field for LCSMO films grown 
on (100) NdGaO$_{3}$.}
\label{rhoH}
\end{figure}

\begin{figure}
\begin{center}
\includegraphics*[width=0.45\textwidth]{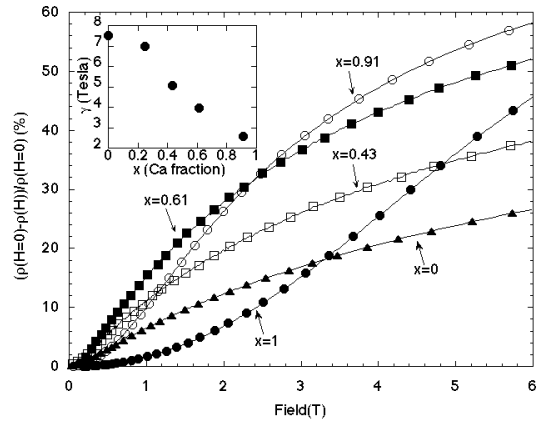}
\end{center}
\caption{Magnetoresistance vs. field at room temperature for the LCSMO films in Fig. 
\ref{rhoH}.  The inset shows the values of $\gamma$ at room 
temperature vs. calcium fraction derived from fitting Eq.\ref{rhoHeq} to the data}
\label{rhoHRT}
\end{figure}

In Figure \ref{rhoH} we show the magnetoresistance at 6 Tesla applied 
field as a function of 
temperature for the films, defined as
\begin{equation}
	MR=\frac{(R(H=0T)-R(H=6T))}{R(H=0T)}.
\end{equation}
For the range of temperatures studied, we see a maximum in the room 
temperature magnetoresistance for the x=0.91 sample, (since T$_{p}$ is 
close to room temperature) however the 
largest magnetoresistance occurs for the pure LCMO sample.  The magnetoresistance at 
77 K for all the samples is linear in applied field, going from $\approx$ 0.5 {\%} for the 
LCMO sample to 3 {\%} for the LSMO sample at 6 Tesla of field.  The field 
dependence of the magnetoresistance at room temperature for the 
samples undergoes a change as would be expected for T$_{p}$ moving 
from above to below room temperature as seen in Fig. \ref{rhoHRT}.  
Near zero field, the curves for the pure LCMO and the x=0.91 sample 
exhibit positive concavity, which is seen for samples with T $>$ 
$T_{p}$, however for higher fields we see that the concavity for the 
x=0.91 sample switches to negative which is that seen for the other 
samples.  As seen in Ref.\cite{Snyder1}, we can fit the change in 
resistance for the case of T$<$ T$_{C}$ to the equation
\begin{equation}
	\rho (H) = \rho_{\infty}+\frac{\Delta}{1+H/\gamma }.
\label{rhoHeq}
\end{equation}
The values of $\gamma$ for samples with x$<$1 at room 
temperature are shown in the 
inset in Fig. \ref{rhoHRT}.  We see that the values of $\gamma$ 
decrease as the Ca fraction increases.  In Ref.\cite{Snyder1} for pure 
LCMO a value of $\gamma$ = 2.7 Tesla was found at 0.9 $T_{C}$, which 
would fit in reasonably into our values, assuming of course that 
$\gamma$ is not strongly temperature dependent.  If $\gamma$ is 
dependent on the relative difference between T and T$_{C}$ or T$_{p}$, we would 
not get the smooth variation seen in the inset of Figure \ref{rhoHRT}, 
since T$_{C}$ and T$_{p}$ are not a monotonic function of x, as seen in Figure 
\ref{rho1}. 

For the pure LCMO sample, we could fit 
the data equally well to the equation proposed in Ref.\cite{Snyder1} 
$\rho(H)= \rho_{\infty}+\Delta/(1+ (H/\beta)^{2})$, or the form 
$\rho(H)= \rho_{0}+aH^{2}+bH^{4}$.  However the use of the first 
equation resulted in values of $\rho_{\infty} < 0$, which is 
unphysical.  The value of $\beta$ is $\approx$ 8.5 T, which is larger 
than that seen in Ref.\cite{Snyder1}, 5.7 T.  The data for the x=0.91 
cannot be fit over the entire range with any of the formulations, 
since it exhibits a concavity change with field.  However, for high 
fields (above 2 Tesla), it can be fit by Eq. \ref{rhoHeq}, giving a 
value of $\gamma$ as seen in Fig. \ref{rhoHRT}.

\section{Conclusions}
We have observed that LCSMO films grow with (100) pseudo-cubic 
orientation on NdGaO$_{3}$ substrates with somewhat rougher surfaces 
and smaller grain size than either pure LCMO or LSMO films.  As the Ca 
fraction decreases, the lattice constant for LCSMO increases towards 
the value for LSMO, resulting in an increase in strain in the system.  
This strain is manifested by a reduction in the Curie temperature, and 
increases in the coercive fields and low temperature resistivity.  We 
have also observed the T$^{2}$ dependence to the resistivity, and 
have observed a correlation between this term and the static term.  
The field dependence to the magnetoresistance for LCSMO films is 
predicted well by the equations in Ref. \cite{Snyder1}, with the 
value of $\gamma$ increasing as the Ca fraction is reduced.

\section{Acknowledgments}
We would like to gratefully acknowledge the assistance of Michael 
Miller for the AFM measurements and Andrew Patton in the 
production of the films.

% \begin{figure}
% \caption{Hysteresis loop at 10 K for pure LSMO and LCMO films grown 
% on (100) NdGaO$_{3}$.}
% \label{Hyst}
% \end{figure}

\end{multicols}
\end{document}